\documentclass[12pt]{article}

\usepackage{amsmath}
\usepackage{amssymb}
\usepackage{amsfonts}
\usepackage{amscd}
\usepackage{amsbsy}
\usepackage{amsthm}
\usepackage{latexsym}
\usepackage{graphicx}
\usepackage{slashed}
\usepackage{color}
\usepackage{cite}
\usepackage{multirow}
\usepackage{tabularx}

\usepackage[utf8]{inputenc}
\usepackage[T1]{fontenc}
\usepackage{mathrsfs}
\usepackage{subfigure}

\usepackage[hyperindex]{hyperref}
\hypersetup{colorlinks=true,linkcolor=blue,citecolor=blue,urlcolor=blue}

\renewcommand{\baselinestretch}{1.2}
\def\beq{\begin{eqnarray}}
\def\eeq{\end{eqnarray}}

\def\ln{\,\mbox{ln}\,}

\def\al{\alpha}
\def\be{\beta}

\def\la{\lambda}

\def\pa{\partial}

\def\si{\sigma}

\def\Ga{\Gamma}

\def\La{\Lambda}

\usepackage{float} 
\usepackage{cite}  
\usepackage{titlesec}

\usepackage{ulem} 

\titleformat*{\section}{\large\bfseries}
\titleformat*{\subsection}{\normalsize\bfseries}

\begin{document}

\begin{center}

{\large\bf
Renormalization group corrections to $\Lambda$CDM model
\\
and observational consequences for $H_0$ tension
}

\vskip 4mm

\textbf{Nicolas R. Bertini$^{a}$} 
\footnote{E-mail address: \ nicolas.bertini@hotmail.com},
\textbf{Marcos H. Novaes$^{b,c}$} 
\footnote{E-mail address: \ marcos.h.pereira@edu.ufes.br},
\textbf{Rodrigo von Marttens$^{c}$} 
\footnote{E-mail address: \ rodrigovonmarttens@gmail.com},
\\
\textbf{and \ \ Ilya L. Shapiro$^{a}$} 
\footnote{E-mail address: \ ilyashapiro2003@ufjf.br}

\vskip 4mm

{\sl $^{a}$ Departamento de F\'{\i}sica, ICE,
Universidade Federal de Juiz de Fora
\\ Campus Universit\'{a}rio, Juiz de Fora, 36036-900, MG, Brazil}
\\[.5ex]

{\sl $^{b}$ PPGCosmo, CCE, Universidade Federal do Espírito Santo
\\
Vitória, 29075-910, ES, Brazil }

{\sl $^{c}$ Instituto de Física, Universidade Federal da Bahia,
40210-340, Salvador, BA, Brazil}
\\[.5ex]

\end{center}

\vskip 4mm

\centerline{\textbf{Abstract}}

\begin{quotation}
\noindent
We explore the renormalization group - based extension of
the $\Lambda$CDM model as a potential solution to the current
cosmological tensions.
In this approach, both the cosmological constant density and Newton’s
constant are allowed to vary with the energy scale, as a consequence of
the remnant effects of massive quantum fields in the low-energy regime.
The corresponding cosmological model is consistent with the principles
of quantum field theory, based on the covariance of the vacuum effective
action, and is characterized by a unique extra parameter $\nu$.
Our analysis yields a best-fit value of
$\nu = - (2.5 \pm 1.3)\times 10^{-4}$, placing the
$\Lambda$CDM limit at the $2\sigma$ region of the $\nu$ posterior.
This narrow range is consistent with data from CMB (Planck), BAO (DESI),
and SN Ia (DES Y5). Our result also alleviates the $H_0$ tension and is
consistent with the previously established constraints from large-scale
structure. In these kind of models there is a link
between cosmology and particle physics.
Our results point to possibility of a new physics, characterized by a mass
spectrum lying below the Planck scale but above the values typically
associated with Grand Unified Theories (GUTs).

\vskip 2mm

\vskip 2mm


\end{quotation}

\section{Introduction}

In the current era of increasingly precise and abundant cosmological
observations, the standard cosmological model is facing what may be
described as an observational crisis. The most enduring and
significant challenge, which already lasts for more than 5 years,
is the so-called Hubble tension. This tension refers to the
discrepancy between the value of Hubble constant $H_0$ inferred
from Cosmic Microwave Background (CMB) observations within the
$\Lambda$CDM framework~\cite{Planck:2018vyg}, and the values
obtained from local distance-ladder measurements based on
Cepheid-calibrated Type Ia supernovae (SNe Ia). This discrepancy has recently reached a level of $\sim 5\sigma$~\cite{Riess:2021jrx}.
Importantly, the CMB-inferred value of $H_0$ is model-dependent,
as it relies on theoretical predictions assuming the $\Lambda$CDM
cosmology, whereas local determinations are model-independent,
being derived directly from cosmographic methods.

Beyond the $H_0$ tension, new results from the DESI DR2 baryon acoustic oscillation (BAO)
analysis suggest a preference for a dynamical dark energy scenario, in contrast with the
cosmological constant of $\Lambda$CDM~\cite{DESI:2025zgx}. When adopting the CPL parameterization
for the dark energy equation of state, the combination of CMB data (Planck 2018), SNe Ia (DESY5),
and BAO (DESI DR2) leads to a tension with $\Lambda$CDM at the level of $4.2\sigma$\footnote{When
considering different SNe Ia surveys, the tension is quantified at $2.8\sigma$
for Pantheon+ and $3.8\sigma$ for Union~3.}.
Moreover, a further discrepancy has been reported in the determination of the matter density
parameter from BAO and SNe Ia data. Specifically, DESI DR2 analyses yield
$\Omega_{\rm m}=0.2975\pm0.0086$, while the DESY5 (SNe Ia) dataset favors
$\Omega_{\rm m}=0.352\pm0.017$. These results disagree at approximately the $3\sigma$ level\footnote{Consistently,
the use of alternative SNe Ia datasets mitigates the tension in the
predictions of the matter density parameter, which become $1.7\sigma$ for Pantheon+ and
$2.1\sigma$ for Union~3.}.

All the results mentioned above have motivated a wide range of proposals
introducing new physics in the late-time sector of $\Lambda$CDM, such
as models with a time-dependent dark energy equation of state
\cite{DeFelice:2012mx, Corasaniti:2002vg, Yang:2018euj}, or even
modifications to the matter sector~\cite{Giani:2024nnv,Giani:2025hhs}.

The mentioned tensions may be addressed by extending the $\La$CDM
model such that there is a slight deviation from it which depends on
time or, equivalently, on the energy scale of the Universe.
The pertinent question is whether the dynamical dark energy scenario
automatically implies the presence of a qualitatively new quantity
such as quintessence or alike, or slightly variable with time
cosmological constant density. From the theoretical side, it is
possible that both cosmological and Newton constants depend on
the energy scale because of the quantum effects. In particular,
this option cannot be ruled out in the framework of semiclassical
gravity, when only matter fields are quantized while the spacetime
metric is classical. The main advantage of this approach is that it
puts together two well-established concepts, such as quantum
matter fields and curved  spacetime. Another advantage is a low
degree of ambiguity owing to the fact that the form of the quantum
effects in question can be defined up to a single free parameter
called $\nu$ \cite{CC-nova}. The $\Lambda$CDM limit is recovered
when $\nu=0$.

In this work, we explore the possibility of extending General
Relativity (GR) introducing an explicit dependence on the energy
scale, such that $G = G(\mu)$ and $\Lambda = \Lambda(\mu)$,
with $\mu$ denoting the renormalization scale. Our approach
assumes the identification of this scale with the Hubble
rate, i.e., $\mu=H$. This hypothesis is based on the qualitative
quantum field theory arguments \cite{CC-nova,Babic2002} and
on the regular scale-identification procedure \cite{Babic:2004ev}
developed for this physical situation. The possibility of the remnant
low-energy (IR) running of the cosmological constant density has
been extensively discussed in these and subsequent works. It was
shown that the IR quantum effects of massive matter fields may
produce the significant effect that could be observed in cosmology.
In short, we can say that the state of the art in quantum field theory (QFT) in curved spacetime does not enable one to either derive these effects or prove that they are impossible \cite{DCCrun} (a very recent discussion
can be found in \cite{FB}).
The reason is that the existing QFT methods are based on the
expansion in the power series in curvature tensor or, more general,
in the derivatives of the metric.
In the next section, we review the corresponding arguments and
some of the publications on this subject.

Assuming that the IR running takes place, in the present work
we demonstrate that this description introduces a significant
modification to the physics of the early Universe.
Phenomenologically, the output is analogous to Early Dark Energy
(EDE) models, while naturally recovering $\Lambda$CDM
dynamics at late times. It is worth mentioning that EDE models
have recently been proposed as a possible solution to the $H_0$
tension~\cite{Poulin:2018cxd,Kamionkowski:2022pkx}. The
similarity with EDE implies
that our model provides a promising framework for such a mechanism,
with the added advantage of being motivated from first principles.

Using the data from Planck 2018 (CMB), DESI DR2 (BAO), and DES Y5 (SN Ia),
we performed a joint statistical analysis for parameter selection. The
results favor our scale-dependent extension over the standard $\Lambda$CDM
scenario, yielding a preferred value of $\nu = - (2.5\pm 1.3)\times10^{-4}$,
with the $\Lambda$CDM limit placed at roughly the $2\sigma$ level.
Moreover, our analysis yields a marginal improvement on the $H_{0}$
tension, with $H_{0}=(68.32\pm0.31)$ km\,s$^{-1}$\,Mpc$^{-1}$ compared to the $\Lambda$CDM prediction
of $H_{0}= (67.34\pm0.30)$ km\,s$^{-1}$\,Mpc$^{-1}$. This slightly larger value is related to the behavior
observed in EDE scenarios, although the effect here is limited
by the small magnitude of the parameter $\nu$.
It is worth emphasizing that our framework introduces only a single additional
parameter, whereas dynamical dark energy models typically require two
additional parameters, and often face difficulties such as phantom crossing.
Also, EDE models usually introduce two or three extra parameters.

This paper is organized as follows. The next section starts with a
brief survey of the derivations of the IR renormalization group
equations for the running cosmological constant density and the
Newton constant. Furthermore, we present the field equations of
general relativity with scale-dependent gravitational and cosmological constants. Sections \ref{sec:background} and \ref{sec:perturbations} are devoted to the background cosmology and scalar perturbations, respectively. Section \ref{sec:observational} addresses the observational analysis. Our conclusions are summarized in Sec.~\ref{sec:conclusions}. Finally, in Appendix~\ref{sec:apdx1} we show that the covariant scale identification, together with the consistency relation, guarantees the conservation of the energy-momentum tensor.
We adopt the following notations and conventions: the Riemann tensor is defined as
$R^{\al}_{\,\,\, \be\si\rho} = \pa_{\si}\Ga^{\al}_{\be\rho} - \pa_{\rho}\Ga^{\al}_{\be\si}+\Ga^{\al}_{\la\si}\Ga^{\la}_{\be\rho}- \Ga^{\al}_{\la\rho}\Ga^{\la}_{\be\si}$, the Ricci tensor as
 $R_{\be\rho}=R^{\al}_{\, \,\,\be\al\rho}$ and the Ricci scalar as
 $R=g^{\al\be}R_{\al\be}$. The metric signature is taken to be $(+,-,-,-)$.

\section{General relativity with running gravitational and
cosmological constants}
\label{sec2}

In this section, we consider the theoretical and phenomenological
backgrounds of our model, i.e., formulate the unique consistent form
of the possible IR running extension of general relativity in the
late Universe.

\subsection{On the
running cosmology in the late Universe}
\label{sec21}

Consider the vacuum quantum effects of massive matter fields in the
low-energy regime, e.g., in the late Universe. In this case, one should
expect a quantum decoupling of massive degrees of freedom and,
consequently, there is no link between divergences and finite part.

Unfortunately, there is no theoretically consistent formal
derivation of the loop contribution to the finite part of the
vacuum effective action in the low-energy regime.
The existing quantum field theory based methods are equivalent
to the expansions in power
series in curvature tensor (or, more general, in metric derivatives,
see, e.g., \cite{OUP}) and this means the expansion around the
flat spacetime metric. However, the problem of interest should
take into account the nonzero cosmological constant, such that the
flat metric is not an appropriate approximation. For this reason,
the consistent derivation of the quantum corrections to the
vacuum effective action is possible only in the fourth-derivative
sectors of the theory \cite{apco} (see \cite{Codello2012,Omar-FF4D}
and \cite{2SimpleQG} for the later discussions). In the two-
and zero-derivative parts, the calculations meet difficulties which
were not resolved, until now.

Behind the desired IR running of the cosmological and Newton
constants, there is an effective action of vacuum, which is
complicated and nonlocal expression, non-polynomial in
curvatures \cite{DCCrun}. How we can distinguish the loop
contributions which can be linked to zero-, two- or higher-derivative
terms in the \textit{classical} action? In cosmology, the answer
should be related to the global scaling of the corresponding
terms. For the background cosmological metric, i.e.,
$g_{\mu\nu} = a^2(\eta)\bar{g}_{\mu\nu}$ (here $\eta$ is the
conformal time), the contributions to the cosmological constant
sector should scale as zero power of $a(\eta)$, the ones to the
Einstein term as $a^{-2}$, the ones to the four-derivative terms
as $a^{-4}$, etc. Beyond these global scaling rules, there are
extra terms with the derivatives of $a(\eta)$, which escape this
classification. However, these terms, including the possible
quantum contributions, represent relatively small corrections
compared to the main scaling laws. It is important that these
arguments are confirmed by direct calculations in the fourth
derivative sector of the theory. The strongest corrections, in
this case, are logarithmic.

Along with this rule of separation in different
sectors, we know that the overall effective action is a covariant
object. Indeed, the covariance does not need to hold for each
sector separately. In the cosmological setting, this implies that
we may expect that the mixing of different sectors will be
manifest when imposing the conservation law. Unfortunately,
only the $\big(1/a^{4}\big)$-type terms admit a consistent evaluation
in the covariant form \cite{apco}.

On the other hand, in the literature, there are different
phenomenological-level derivations of the remnant renormalization
group running of the effective energy density of the vacuum. The
first such calculation \cite{CC-nova} was based on the hypothesis
of the  quadratic decoupling in the IR. This is the standard form of
the low-energy decoupling, associated with the Appelquist and
Carazzone theorem \cite{AC}. In the mentioned works, starting
from \cite{apco}, the  gravitational analog of this
theorem was proved, but only in the four-derivative sector of the action.
Thus, this is a natural, albeit not verified expectation to meet the
same form of decoupling for the cosmological constant density.
Since the beta function for the cosmological constant density
based on the Minimal Subtraction scheme  is proportional to the
fourth power of masses of the quantum fields, the quadratic
decoupling gives the contribution proportional to the linear
combination of the squares of these masses.
Thus, we arrive at the energy density of vacuum that behaves as
\beq
\rho_\La(\mu)\,=\, \rho_\La (\mu_0) \,+\,
\frac{3\nu}{8\pi G}\big(\mu^2 - \mu_0^2\big),
\label{rhoCCmu}
\eeq
where $G$ is the Newton constant and $3\nu/8\pi G$ is
proportional to the squares of the masses.
At this point, one can formulate
the two critical questions. The first one is the identification of the
energy scale $\mu$. The answer to this question has been given in
\cite{CC-nova,Babic:2004ev}. After discussing various options, the
most natural one, in the cosmological setting, is the Hubble
parameter, $\mu = H$. The second problem is to complement the rule (\ref{rhoCCmu}) with an appropriate conservation law. Only
with this completion one can arrive at the closed system of equations
and construct the cosmological model. The first such model
\cite{CC-PLB} assumed the energy exchange between vacuum
and matter. However, shortly after this, there was a criticism of
this approach in Refs.~\cite{WangMeng,OpherPelin}. The point
noted in these papers was that significant energy can go from
vacuum to matter only in the form of the creation of massive particles.
However, such creation is possible only in the
early Universe since it requires the energies of the gravitational
quanta to be higher than the threshold of the masses of particles.

On the other hand, there is even a simpler argument against the
original model \cite{CC-PLB} and all subsequent developments
taking place in the last decades. As we already stressed above, the
quantum
field theory arguments indicate that the effective action of vacuum
is a covariant object on its own, without the matter sector. Thus,
one has to provide the fulfillment of the conservation law which
does not involve the matter filling the Universe.\footnote{This
may not be true in the very early epochs, e.g., owing to the
supersymmetry that might mix matter and gravity.}

An interesting detail about the covariance and its implication
for the conservation law is that the fourth- and higher-derivative
sectors of the theory
are Planck-suppressed after the very early epochs of the
evolution of the Universe, at least after the end of inflation
(even assuming it is described by the Starobinsky model \cite{star})
and reheating periods. Thus, one has to rely on the energy exchange
between the cosmological constant
(i.e., $\mathcal{O}(a^0)$)
and the Einstein
(i.e., $\mathcal{O}(1/a^2)$)
sectors of the vacuum
effective action. The corresponding model has been constructed in
\cite{Shapiro:2004ch} and explored at the level of cosmic
perturbations in \cite{Fabris2010}, using the general methodology
developed earlier in \cite{Fabris2006}. The first output of the
described conservation law is the running described below in
Eq.~(\ref{eq:runG}). In what follows, we reproduce and recalculate
in a partially new way the main elements of the model of
\cite{Shapiro:2004ch} and
\cite{Fabris2010}. After this, we consider applications of this
model to the new cosmological data and find that it is quite
operational and promising in their consistent explanation.

\subsection{Formulation of the running cosmology model}
\label{sec22}

Consider that $G$ and $\Lambda$ may run with the energy
scale $\mu$. Then the modified Einstein equations read
\begin{align}
R_{\al\be} - \frac{1}{2}g_{\al\be}R
\,=\, 8\pi G(\mu)\big[T_{\al\be}
\,+\, g_{\al\be} \rho_{\Lambda}(\mu)  \big]\,.
\label{eq:FE}
\end{align}
Here $T_{\al\be}$ is the energy-momentum tensor of matter
and radiation, and $\rho_{\Lambda}=\Lambda/(8\pi G)$. We assume
that $T^{\al\be}$ takes the form of a perfect fluid with energy density
$\rho$, pressure $p$ and four-velocity $u^{\al}$, i.e,
\begin{align}
T^{\al\be} = (\rho+p)u^{\al}u^{\be}-g^{\al\be}p\,.\label{eq:EMT}
\end{align}

Within this framework, a suitable scale-setting procedure must be
adopted to relate $\mu$ to relevant physical quantities. Regardless
of the corresponding choice, the consistency with general covariance requires $\mu$ to be treated as a scalar quantity defined on spacetime, rather than as a fixed external parameter. Otherwise, the quantities $G(\mu)$ and $\rho_\Lambda(\mu)$ would not transform covariantly under coordinate transformations. In this sense, $\mu$ plays the role of an effective scalar field encoding the local physical scale associated with the gravitational system.

Once $G(\mu)$ and $\rho_\Lambda$ acquire spacetime dependence through $\mu$, the Bianchi identities no longer imply the standard conservation law automatically. Indeed, taking the divergence of Eq.~\eqref{eq:FE} yields
\begin{align}
\nabla_{\al}T^{\al}_{\;\;\be} = -\nabla_\be \rho_\Lambda - (T^{\al}_{\;\;\be} + \rho_\Lambda \delta^{\al}_{\be})G^{-1}\nabla_{\al}G\,,
\end{align}
showing that, in general, $\nabla_{\al}T^{\al\be} \neq 0$. However, as shown in Appendix~\ref{sec:apdx1}, under consistent assumptions in
cosmology, the conservation law $\nabla_{\al}T^{\al\be} = 0$
holds provided that
\begin{align}
\frac{d\rho_{\Lambda}}{d\mu}
- \frac{3\mu^{2}}{8\pi}\frac{dG^{-1}}{d\mu}\,=\,0 ,
\label{eq:consist}
\end{align}
with $\mu^{2}:=\frac{1}{3}u^{\al}u^{\be}G_{\al\be}$, where $G_{\al\be} = R_{\al\be}-g_{\al\be}R/2$ is the Einstein tensor.

Notably, in the context of background cosmology, this identification provides  $\mu = H$ (details in the next section), in agreement with the most well-motivated identification obtained from RG-based scale-setting procedures \cite{Babic:2004ev, Shapiro:2004ch, Domazet:2010bk}, as well as the approach of \cite{Bertini:2024naa}.

A consistent identification of $\mu$ as a scalar field is crucial, as it allows the extension of the framework to more general settings. Although the identification above arises naturally from the field equations, it is not unique --- other definitions may also lead to $\mu = H$ (see, e.g., \cite{Fabris2010}) and reproduce the same background behavior. Nevertheless, this choice can affect the resulting dynamics and yield different predictions depending on the physical system, particularly at the level of cosmological perturbations.
As will be shown in Sec.~\ref{sec:perturbations}, this ambiguity does not arise if the scalar field associated with $\mu$ is compatible with the conservation law $\nabla_{\al}T^{\al\be}=0$, since this condition ensures that the perturbation of $\mu$ does not appear in the linearized field equations.

Eq.~\eqref{eq:consist} implies that the specific functional forms of $\rho_\Lambda(\mu)$ and $G(\mu)$ cannot be arbitrary. Typically, one function is derived from the RG equations via the corresponding $\be$-functions, $\be_{X} = \mu\,dX/d\mu$, where $X$ can be $G$ or $\rho_\Lambda$, while the other is obtained by solving the consistency relation \eqref{eq:consist}. Based on well-motivated dimensional arguments \cite{Farina:2011me} (see also \cite{Reuter:2003ca, Nelson:1982kt, Fradkin:1981iu}), we assume that the infrared $\be$-function of $G^{-1}$ is given by $\be_{G^{-1}} = 2\nu M_\text{Planck}^{2}$, where $\nu$ is a dimensionless constant, the running of $G$ reads
\begin{align}
G(\mu) = \frac{G_0}{1+\nu\ln(\mu^{2}/\mu_0^2)},
\label{eq:runG}
\end{align}
where $G_{0}\equiv M_\text{Planck}^{-2}$ and $\mu_0$  is the reference scale at which $G(\mu_0) = G_0$. Integrating \eqref{eq:consist}, the corresponding running of $\rho_{\Lambda}$ becomes
\begin{align}
\rho_{\Lambda}(\mu)
\,=\, \rho_{\Lambda 0} + \frac{3\nu}{8\pi G_0}(\mu^{2}-\mu_0^2),
\label{runCCmu0}
\end{align}
where $\rho_{\Lambda0} = \rho_{\Lambda}(\mu_0)$.
Compared to (\ref{rhoCCmu}), we just fixed $G\equiv G_0$
in the \textit{r.h.s.}.

Just to finish the discussion of our model, let us mention two
more theoretical aspects.

\textit{i)}\ Expression (\ref{rhoCCmu}) with the identification
of scale $\mu = H$ is a direct consequence of the covariance of
effective action. Since there is a cosmological constant term and
the quantum contributions represent relatively small corrections
to this term, the effective action of vacuum admits the expansion
into power series in the derivatives of the metric. Owing to
the covariance, such expansion consists only from the even
terms \cite{CC-nova,PoImpo}. In the cosmological metric case,
(\ref{rhoCCmu}) is nothing but the possible first term of such
an expansion.

\textit{ii)} \ One can arrive at the running (\ref{eq:runG})
independently, using covariance and dimensional arguments
\cite{Farina:2011me}. For this end, it is sufficient to note that
the dimension of $1/G$ is the square of mass. This means, the
beta function for $1/G$ should be a bilinear combination
of the masses $m_i$ of matter fields,
\beq
\mu\, \frac{d}{d\mu} \,
\frac{1}{G}
\,\,=\,\, \sum_{i,j}A_{ij}\,m_i \,m_j \,,
\label{rengrG}
\eeq
where the coefficients $A_{ij}$ are constants at the one-loop
level and may depend on the coupling constants of the
underlying quantum theory in higher loops. It is easy to see
that the solution of this equation is (\ref{eq:runG}), where
$\nu$ is a phenomenological parameter related to the mass
scale of the quantum theory of matter fields.

Starting from (\ref{eq:runG}) and using the conservation equation
(\ref{eq:consist}) in the opposite direction, we arrive back to
Eq.~(\ref{rhoCCmu}). Thus, assuming that there is no
energy exchange between vacuum and matter, we arrive at
the unique possible model consisting of (\ref{eq:runG})
and (\ref{runCCmu0}). The unique free constant of this
model is the parameter $\nu$ which roughly equals to the
ratio between the linear combination of the squares of the
heaviest masses of the underlying quantum theory and the
square of the Planck mass. Such a linear combination may
be dominated by bosons, which means $\nu > 0$, or fermions,
with $\nu < 0$. The magnitude of $\nu$ may be extremely
small for the Minimal Standard Model and be of the order
of $10^{-6}$ or even larger in the Grant Unification Theories
with supersymmetry and high multiplicities of the quantum
fields.  The limits on $\nu$ from the power spectrum of the
density perturbations, LSS data (of the year 2010) and
the $F$-test are not very restrictive, with $|\nu| < 0.001$.
This value indicates that the underlying quantum theory
may be of, basically, any origin -- except involving the
particles with the masses of the Planck order of magnitude.
In what follows, we shall check what are the bounds on
$\nu$ obtained from other observables.

\section{Background cosmology equations}\label{sec:background}

We assume a homogeneous and isotropic Universe, as given by the Friedmann–Lemaître–\\Robertson–Walker (FLRW) line element
\begin{equation}
	ds^{2} = dt^{2} - a^{2}(t)\delta_{ij}dx^i dx^j\,,\label{eq:Mbackground}
\end{equation}
where $a$ is the scale factor. This implies that $\mu^2 =  H^{2}$, where $H=\dot{a}/a$ is the Hubble parameter, and the two independent components of Eq.~\eqref{eq:FE} are
\begin{align}
	H^{2} = \frac{8\pi G(H)}{3}\big[ \rho + \rho_{\Lambda}(H) \big]\,,\label{eq:fried1}\\
	2\dot H + 3H^{2} =  - 8\pi G(H) \big[  p - \rho_\Lambda(H) \big]\,.\label{eq:fried2}
\end{align}
Using the running laws for $G(H)$ and $\rho_\Lambda(H)$ introduced in the previous section, these equations are equivalent to the standard continuity equation
\begin{equation}
	\dot\rho +3H(\rho + p) = 0\,, \label{eq:cont}
\end{equation}
which is also obtained directly from $\nabla_{\al}T^{\al\be}=0$.

To analyze the phenomenological implications of this scenario, one needs to specify the value of the reference scale $\mu_0$. Since RG effects are expected to be more significant in the early Universe, we follow the approach of \cite{Bertini:2024onw, Bertini:2024naa} and assume that the future de Sitter phase is free of such corrections. Then, evaluating $u^{\al}u^{\be}G_{\al\be}$ with the de-Sitter metric, one gets $\mu_0 = \sqrt{\Lambda_0/3}$.

The above system of equations can be solved analytically to express $H(a;\nu)$. To this end, it is important to note that the parameter $\nu$ is expected to satisfy $|\nu| \ll 1$. From the theoretical perspective, $\nu$ is proportional to the ratio between the squared masses of the underlying quantum fields and the Planck mass squared, and is therefore naturally suppressed \cite{Shapiro:2004ch}. Moreover, previous analyses based on large-scale structure and CMB observations constrain $|\nu|\lesssim 10^{-3}$ \cite{Shapiro:2004ch}. Therefore, one may expand the running law \eqref{eq:runG} in the form $G^{-1}(\mu) = G_0^{-1}(\mu/\mu_0)^{2\nu} = G_0^{-1} (1+\nu \ln \mu^{2}/\mu_0^2) + {\cal O}(\nu^2)$ and then obtain the corresponding $\rho_{\Lambda}(\mu)$ by solving Eq.~\eqref{eq:consist}, which gives
\begin{align}
\rho_\Lambda (\mu) = \rho_{\Lambda 0} +\frac{3\nu}{8\pi (1+\nu)}(\mu^2 G^{-1} - \mu_0^2 G_{0}^{-1})\,.
\end{align}
Next, Eq.~\eqref{eq:cont} can be solved separately for radiation and matter, yielding $\rho_\text{r} = \rho_\text{r0}/a^{4}$ and $\rho_\text{m} = \rho_\text{m0}/a^3$, respectively, where $\rho_\text{r0}$ and $\rho_\text{m0}$ are integration constants. Substituting these expressions into Eq.~\eqref{eq:fried1}, we obtain
\begin{align}
\left( \frac{H}{H_0} \right)^{2(1+\nu)} = \Omega^{\nu}_{\Lambda 0}\Bigg[ (1+\nu)\left( \frac{\Omega_\text{r0}}{a^{4}} + \frac{\Omega_\text{m0}}{a^{3}}+\Omega_{\Lambda 0}\right) - \nu \Omega_{\Lambda 0} \Bigg]\label{eq:Hnu}
\end{align}
where $H_0$ is the Hubble parameter evaluated today ($a=1$) and $\Omega_{x0}:=8\pi G_0 \rho_{x0}/3H_0^2$, where $x$ denotes radiation, matter or cosmological constant. Solving the closure relation for $\Omega_\mathrm{m0}$ yields
\begin{align}
\Omega_\text{m0} = \Omega_{\Lambda 0}(\nu - 1) - \Omega_\text{r0} + \frac{\Omega^{-\nu}_{\Lambda 0}}{1-\nu} = 1 - \Omega_\text{r0} - \Omega_{\Lambda 0} + (1-\Omega_{\Lambda 0} - \ln\Omega_{\Lambda 0})\nu + {\cal O}(\nu^2).
\end{align}
Finally, substituting this result into \eqref{eq:Hnu} and expanding to first order in $\nu$, one gets 
\begin{align}
	\frac{H}{H_0}= F\Bigg\{1- \frac{\nu}{2}\Bigg[ \frac{1}{F^{2}}\Bigg( \frac{\Omega_{r0}+\ln\Omega_{\Lambda0}}{a^{3}} - \frac{\Omega_{r0}}{a^{4}}\Bigg)+\ln \frac{F^{2}}{\Omega_{\Lambda0}} \Bigg]  \Bigg\}\,,
\end{align}
where $F^2 = \Omega_{\Lambda 0} + \Omega_\text{r0}a^{-4} + (1 -\Omega_{\Lambda 0} - \Omega_\text{r0}) a^{-3} $.

\begin{figure}[h!]
    \centering
    \includegraphics[width=0.59\textwidth]{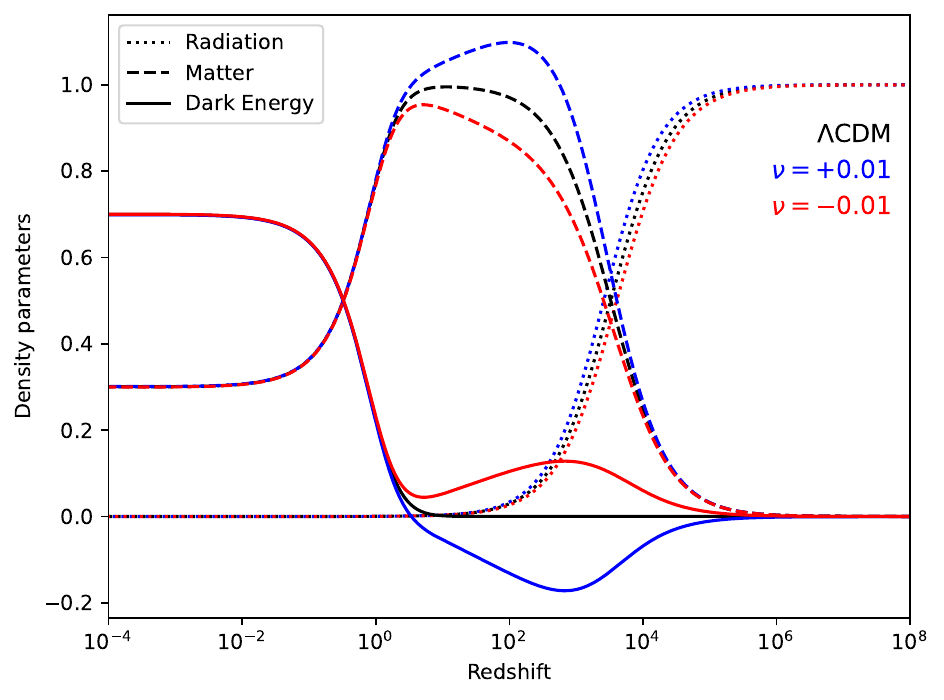}
    \caption{Time-dependent density parameters $\Omega_{x}(z) \equiv \rho_{x}/\rho_\text{crit}$, as a function of redshift, for $\nu = \pm 0.01$.}
    \label{fig:omegas_ratio}
\end{figure}

In order to illustrate the impact of the parameter $\nu$ on the background evolution,
we show in Fig.~\ref{fig:omegas_ratio} the evolution of the time dependent density parameters as a function of redshift $\Omega_{x}(z) \equiv \rho_{x}/\rho_\text{crit}$, where $\rho_\text{crit} = 3H^2/(8\pi G)$. This figure makes clear that our proposed extension to GR behaves
as an early-time deviation from $\Lambda$CDM. For positive values of $\nu$, the
contribution of the dark energy component is negative, while the deviation is positive
for $\nu < 0$. At late times, however, the model smoothly converges to the standard
$\Lambda$CDM behavior. In Fig.~\ref{fig:omegas_ratio} we display the cases with $|\nu|=0.01$,
which is admittedly larger than the observationally allowed values (as will be discussed later),
but it serves to clearly illustrate the role of $\nu$\footnote{For values of $\nu$ within
the characteristic order of the constraints imposed by the CMB ($\nu\sim 10^{-4}$),
the deviations in the background dynamics are at the level of $1\%$.}.

Since the parameter $\nu$ modifies the dynamics only at early times, it is not expected to
have a direct impact on Type Ia supernova observations, although indirect effects may arise
through parameter correlations. On the other hand, its influence should be more pronounced
in BAO measurements, through changes in the sound horizon, and in CMB physics at the epoch
of last scattering. Remarkably, as can be seen in Fig.~\ref{fig:omegas_ratio}, negative
values of $\nu$ lead to an increase in the $\Lambda$ component at early times, which is a
characteristic feature of EDE models. In that sense, our model provides a
physical mechanism for such scenarios, with the added advantage of being motivated from
first principles. As it is well-known, EDE models can alleviate the $H_0$
tension~\cite{Poulin:2018cxd,Kamionkowski:2022pkx}, our framework offers a promising
avenue in this regard.

\section{Scalar perturbations}\label{sec:perturbations}

Introducing the perturbations in the metric, $g_{\al\be} \rightarrow g_{\al\be}+ \delta g_{\al\be}$, the linearized form of Eq.~\eqref{eq:FE} can be written as
\begin{align}
    \delta G^{\al}_{~\be} =8\pi G(\mu) M^{\al}_{~\be}\, \delta\mu + 8\pi G(\mu) \delta T^{\al}_{~\be}.
\end{align}
where $\delta\mu$ is the perturbation of $\mu$ and
\begin{align}
    M^{\al}_{~\be}:=\frac{d\rho_{\Lambda}}{d\mu}\delta^{\al}_{\be}  - \frac{G^{\al}_{~\be}}{8\pi}\frac{dG^{-1}}{d\mu}\,.
\end{align}
Although $\delta\mu$ can be computed directly (see \cite{Bertini:2024naa}), this is not necessary, since its contribution vanishes from the linearized field equations. This result can be obtained by taking the divergence of the field equations \eqref{eq:FE} and use $\nabla_{\al}\rho_\Lambda = (\nabla_{\al}\mu )d\rho_\Lambda/d\mu$ and $\nabla_{\al}G = (\nabla_{\al}\mu) dG/d\mu$, which leads to $\nabla_{\al}T^{\al}_{~\be} = M^{\al}_{~\be}\nabla_{\al}\mu$. Solving $M^{\al}_{~\be}$, one gets
\begin{align}
    M^{\al}_{~\be}  = \frac{\nabla^{\al}\mu}{\nabla^{\si}\mu \nabla_{\si}\mu}\nabla_{\la}T^{\la}_{~\be}\,,
\end{align}
As shown in Appendix~\ref{sec:apdx1}, $\nabla_{\la}T^{\la}_{~\be} = 0$. Consequently, the linearized field equations reduce to
\begin{align}
    \delta G^{\al}_{~\be} = 8\pi G(\mu) \delta T^{\al}_{~\be}. \label{eq:FElin}
\end{align}

To analyze the consequences of this result, we consider scalar modes. Assuming a spatially flat background with line element \eqref{eq:Mbackground}, we introduce the perturbed metric in the Newtonian gauge
\begin{align}
    ds^2 = (1+2\Psi)dt^{2} - a^2(t)(1-2\Phi)\delta_{ij}dx^i dx^j\,.
\end{align}

Equation \eqref{eq:FElin} shows that the linearized field equations have the same structure as in standard GR, i.e.,
\begin{align}
\frac{2}{a^{2}}\nabla^{2}\Phi - 6H( \dot{\Phi} + H\Psi) = 8\pi G(H) \delta T^{0}_{\,\,\,0}\,,\label{eq:lin-00}
\end{align}
\begin{align}
\pa_{i}(\dot{\Phi} + H\Psi) = 4\pi G(H)\delta T^{0}_{\,\,\,i}\,, \label{eq:lin-0i}
\end{align}
\begin{align}
\bigg[\ddot{\Phi} + H(3\dot{\Phi}+\dot{\Psi}) + (3H^{2}+2\dot{H})\Psi - \frac{1}{2a^{2}}\nabla^{2}(\Phi-\Psi)\bigg]\delta^{i}_{j}
\nonumber\\
+ \frac{1}{2a^{2}} \pa^{i}\pa_{j}(\Phi - \Psi) = -4\pi G(H)\delta T^{i}_{\,\,\, j}, \label{eq:lin-ij}
\end{align}
but here the gravitational coupling is scale dependent, $G(H)$ and $H=H(t;\nu)$. The perturbed energy-momentum tensor components are found from \eqref{eq:EMT}. At first order, the four-velocity reads $u^{\al} = (1-\Psi,\vartheta^i)$, which leads to $\delta T^{0}_0 = \delta\rho$, $\delta T^{0}_{i} = -a^2(\rho+p)\delta_{ij}\vartheta^i$ and $\delta T^{i}_{j} = -\delta^i_j\, \delta p$,  where $\delta\rho$ and $\delta p$ are the energy density and pressure perturbations, respectively. From the off-diagonal part of Eq.~\eqref{eq:lin-ij} one finds
\begin{align}
    \Phi = \Psi\,,
\end{align}
so that the slip parameter $\eta \equiv \Phi/\Psi = 1$, as in standard GR.

Into Fourier space, Eqs. \eqref{eq:lin-00}-\eqref{eq:lin-ij} can be written as
\begin{align}
\frac{k^{2}}{a^{2}}\Phi + 3H( \dot{\Phi} + H\Phi) = - 4\pi G(H) \delta\rho\,,\label{eq:lin-002}
\end{align}
\begin{align}
k^{2}(\dot{\Phi} + H\Phi) =  4\pi  G(H)a^{2}(\rho+p)\theta\,, \label{eq:lin-0i2}
\end{align}
\begin{align}
\ddot{\Phi} + 4H\dot{\Phi} + (3H^{2}+2\dot{H})\Phi  = 4\pi G(H) \delta p \, . \label{eq:lin-ij2}
\end{align}
where $\theta\equiv \pa_{i}\vartheta^{i}$ denotes the velocity divergence.

\section{Data analysis and methodology}\label{sec:observational}

To test the viability of introducing the energy scale $\mu$, we perform a Bayesian statistical analysis based on the most recent public available cosmological data. For the parameter inference we employ the \texttt{Cobaya} code~\cite{Torrado:2020dgo} for a Markov Chain Monte Carlo (MCMC) exploration of the parameter space. We run multiple independent chains and impose the Gelman–Rubin convergence criterion $R < 0.01$, ensuring that the sampled distributions are stable and statistically robust. Our aim is to determine whether the introduction of a scale-dependent framework can improve the consistency between different cosmological probes compared to the standard $\Lambda$CDM scenario.

\subsection{Datasets}

In our statistical analysis we combine the following complementary probes of the expansion history and structure formation, spanning both early- and late-time cosmology:

\begin{itemize}
    \item \textbf{Cosmic Microwave Background (CMB):} We include the full set of Planck 2018 likelihoods,
    which provide the most precise measurements of the CMB anisotropies to date. At low multipoles,
    we use the standard temperature and polarization likelihoods. At high multipoles, we adopt the Planck
    NPIPE CamSpec likelihood, which delivers improved systematics control in the TT, TE, and EE
    spectra~\cite{Efstathiou:2019mdh,Rosenberg:2022sdy}. Finally, we also include the CMB lensing likelihood
    from Planck PR4 \cite{Carron:2022eyg}\footnote{Based on the GitHub repository \texttt{carronj/planck\_PR4\_lensing}.}.
    The lensing signal is sensitive to the integrated distribution of matter and provides an important cross-check on late-time growth.

    \item \textbf{Baryon Acoustic Oscillations (BAO):} We use the measurements of baryon acoustic oscillations from the Data Release 1 (DR1) of the Dark Energy Spectroscopic Instrument (DESI)~\cite{DESI:2024mwx}. DESI DR1 includes over 6 million galaxy and quasar redshifts, providing the most precise BAO distance ladder to date in the redshift range $0.1 < z < 2.1$. The BAO feature serves as a standard ruler, allowing us to probe the comoving angular diameter distance $D_M(z)$ and line-of-sight distance $D_H(z)\equiv c/H(z)$ at multiple redshifts.

    \item \textbf{Type Ia Supernovae (SN Ia):} As a late-time probe of the expansion history, we include the DESY5 SN Ia sample \cite{DES:2024jxu}. DESY5 contains more than 1500 spectroscopically confirmed Type Ia supernovae, uniformly observed and reduced, spanning the redshift range $0.01 < z < 1.2$. Compared to earlier compilations, DESY5 offers improved control of calibration and selection effects, since all of its high-redshift supernovae were observed with the same instrument, ensuring a homogeneous dataset. The cosmological observable extracted from these measurements is the luminosity distance as a function of redshift, $d_L(z)$, typically expressed through the distance modulus $\mu = m - M$, where $m$ and $M$ are the apparent magnitude and the absolute magnitude of a given SN Ia. This quantity is particularly sensitive to the late-time acceleration, but it provides an essential complement to CMB and BAO data, helping to break degeneracies between certain cosmological parameters.

    \item \textbf{$H_0$ prior:} Given the potential of our model to alleviate the $H_{0}$ tension, we also examine an additional scenario in which we include a gaussian prior on the Hubble constant, $H_{0} = (73.0 \pm 1.4)$ km\,s$^{-1}$\,Mpc$^{-1}$~\cite{Riess:2020fzl}.
\end{itemize}

\subsection{Results}

\begin{figure}[ht]
    \centering
    \includegraphics[width=0.48\textwidth]{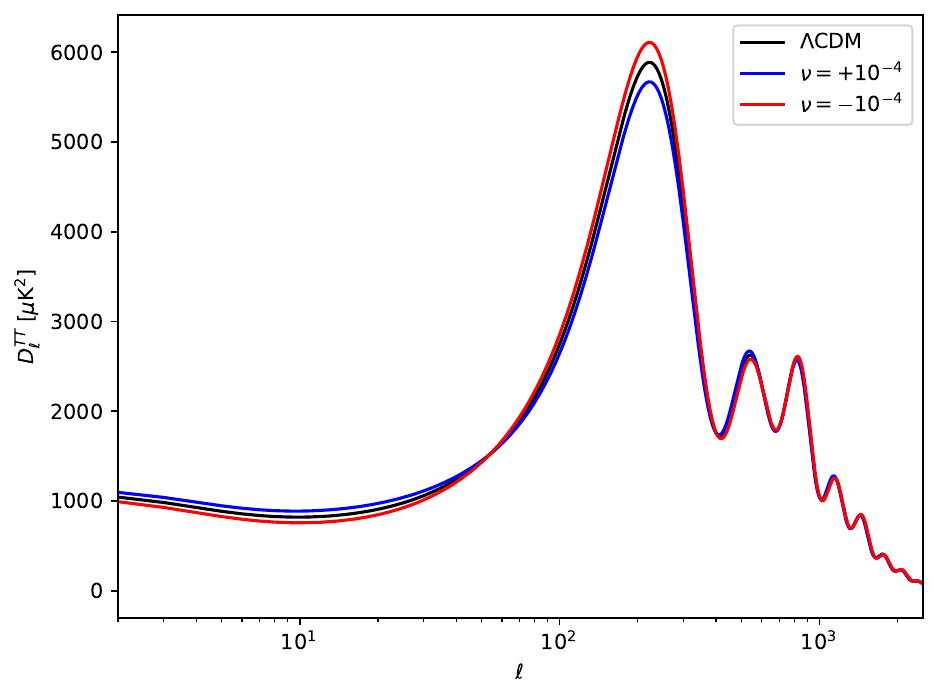}
    \hfill
    \includegraphics[width=0.48\textwidth]{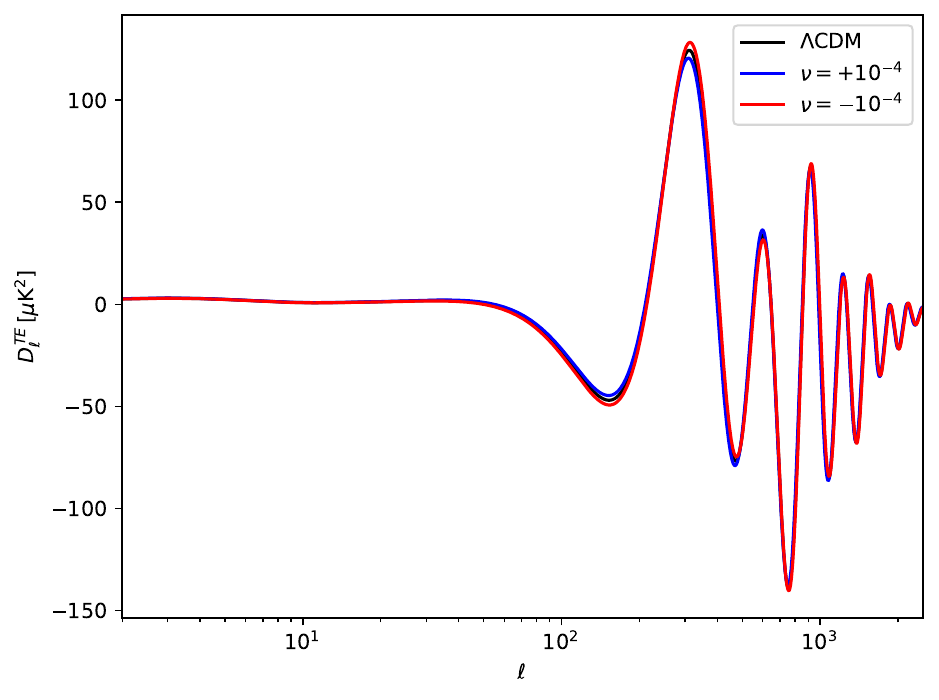}
    \caption{CMB angular power spectra for the scale-dependent model compared to $\Lambda$CDM.
    Both TT (left) and TE (right) spectra are shown for {\color{blue}$\nu = \pm 10^{-4}$}, illustrating
    the effect of the running parameter on early-time physics.}
    \label{fig:CMB_spectra}
\end{figure}

To evaluate the performance of our model against observations, we carry out a Bayesian
parameter inference analysis. Since the background probes are mainly sensitive to low-$z$
physics, while the main features of our model manifest at early times, we restrict our
study to dataset combinations that include CMB measurements. This choice is further
motivated by the strong sensitivity of the CMB spectra to the parameter $\nu$, since it
affects the underlying physics at the epoch of recombination. To illustrate this effect,
Fig.~\ref{fig:CMB_spectra} displays the temperature spectrum and the temperature–polarization
cross-correlation for $|\nu|=10^{-4}$, compared to the standard $\Lambda$CDM case.

Specifically, in our statistical analysis we consider
the following cases: ($i$) CMB only; ($ii$) CMB + $H_0$ prior; ($iii$) CMB + BAO + SN Ia;
and ($iv$) CMB + BAO + SN Ia + $H_0$ prior. Moreover, for our parameter selection we use the following set of free parameters:
six standard $\Lambda$CDM parameters, namely the physical baryon density
$\omega_{\rm b} \equiv \Omega_{\rm b} h^2$, the $\Lambda$ density parameter
$\Omega_{\Lambda}$\footnote{We replace the physical CDM density
$\omega_{\rm c} \equiv \Omega_{\rm c} h^2$ with $\Omega_\Lambda$ as our choice of parameterization.},
the Hubble constant $H_0$, the reionization optical depth $\tau_{\rm reio}$,
the amplitude of the primordial scalar power spectrum $A_s$ and its spectral index $n_s$; plus the new scale-dependent
parameter $\nu$. For all parameters we have used flat widely conservative priors.

\begin{figure}[ht]
    \centering
    \includegraphics[width=0.45\textwidth]{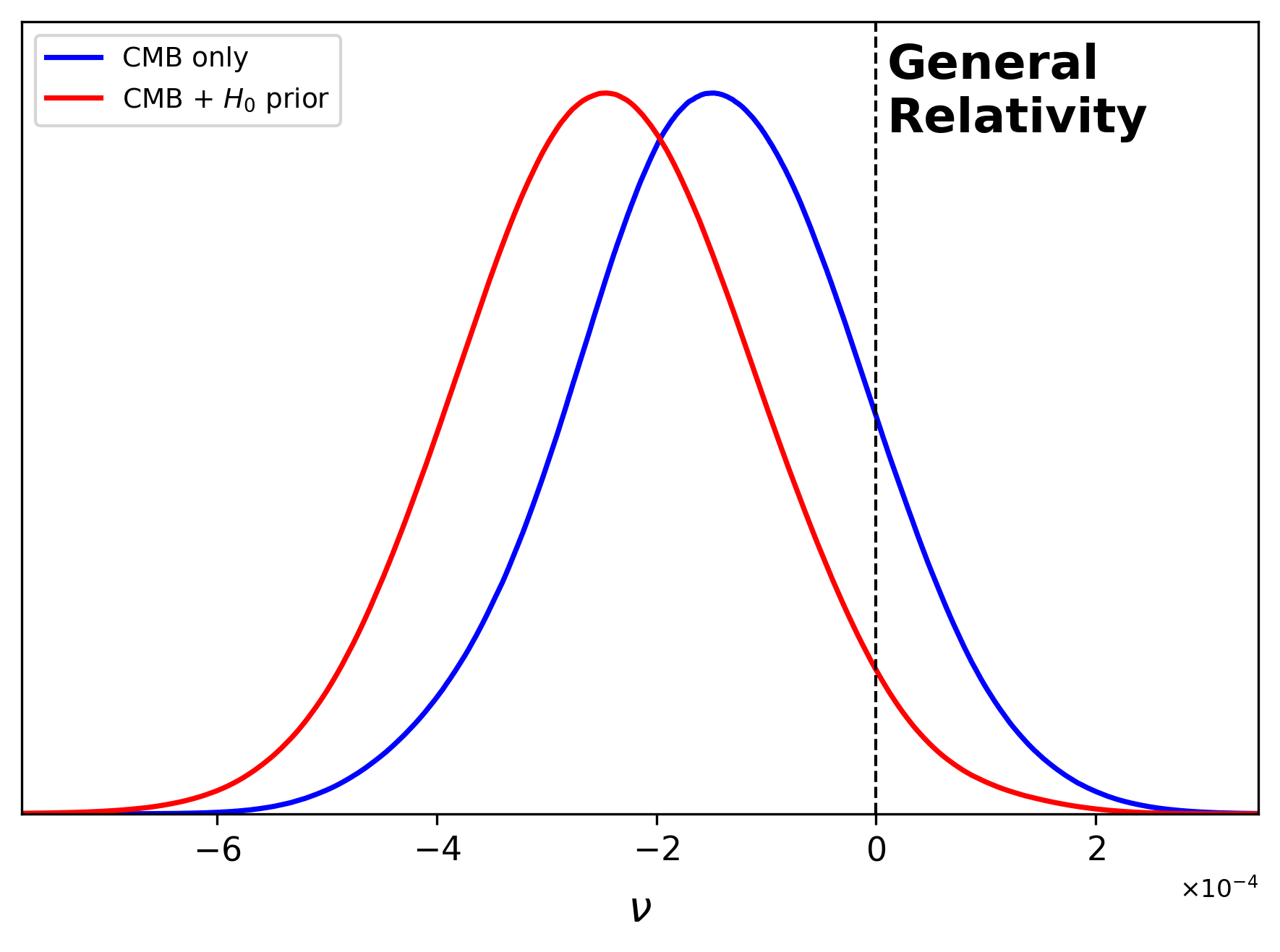}
    \includegraphics[width=0.45\textwidth]{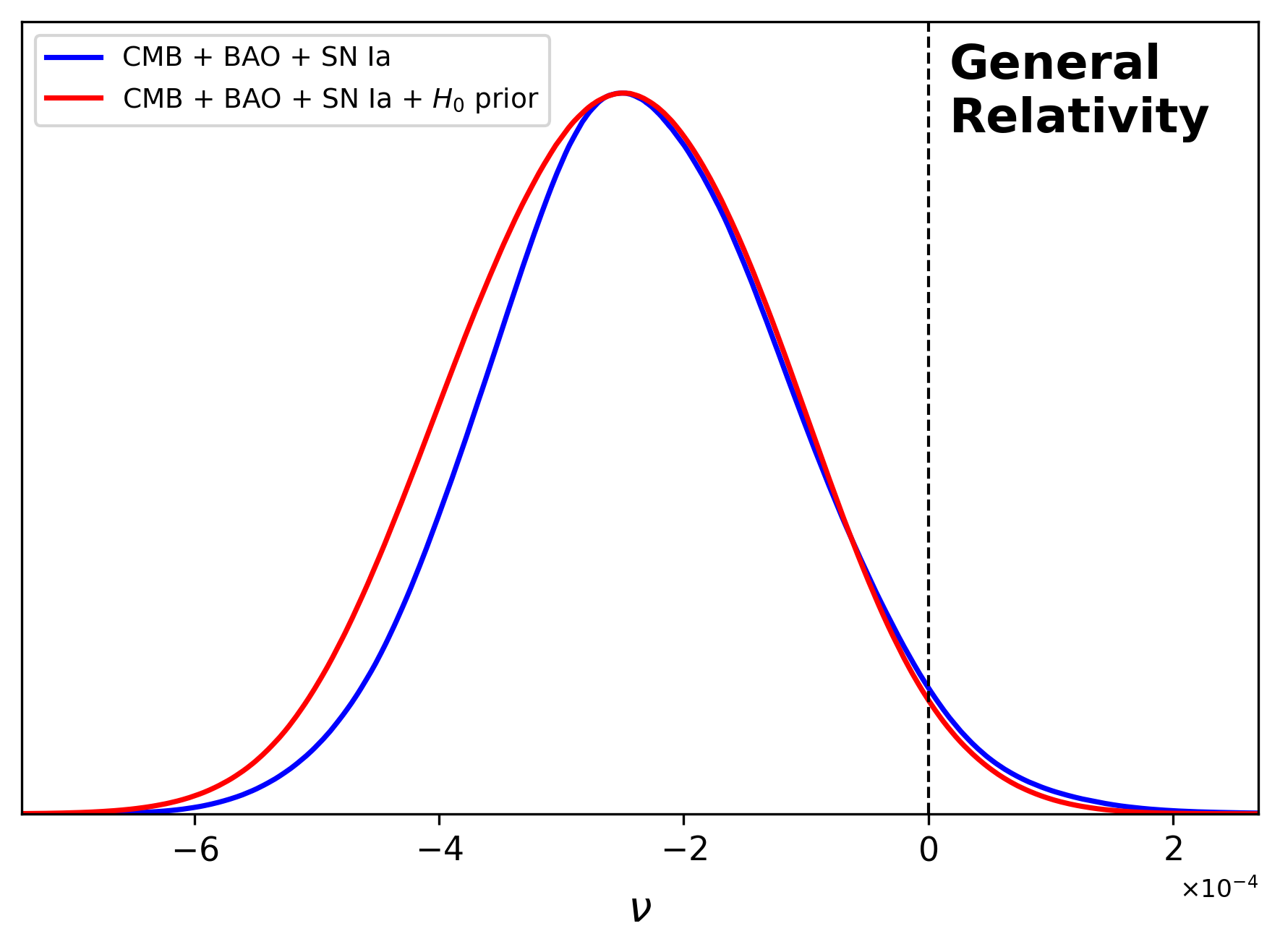}
    \caption{Marginalized posterior distributions for the parameter $\nu$. The dashed vertical line indicates the GR limit, $\nu=0$. \textbf{Left panel: }results from CMB only and CMB + $H_0$ prior. \textbf{Right panel: }results from CMB + BAO + SN Ia and CMB + BAO + SN Ia + $H_0$ prior.}
    \label{fig:nu_posteriors}
\end{figure}

The results of the statistical analysis across the different dataset combinations are summarized in Table~\ref{tab:cosmo_results}.
First, the constraints on the parameter $\nu$ consistently favor negative values of the order of $10^{-4}$ across all dataset combinations,
with the statistical significance ranging from a mild indication to nearly the $2\sigma$ level. For the CMB only case, the result
$\nu=-0.00014\pm0.00013$ corresponds to a $1.1\sigma$ deviation from the $\Lambda$CDM limit, which can be understood as a statistical fluctuation.
When the $H_0$ prior is included, the deviation increases to $1.7\sigma$, indicating that both parameters are indeed correlated, as expected because of the
previously mentioned EDE-like behavior of the model. When BAO and SN Ia data are added to CMB, the preference for $\nu<0$ becomes more pronounced,
with $\nu=-0.00023\pm0.00012$, which represents a $1.9\sigma$ deviation. This result indicates that the inclusion of late-time background information
enhances the preference for $\nu<0$. Finally, when $H_0$ prior is also included in this case, the deviation from $\Lambda$CDM remains unaffected as
$1.9\sigma$, with $\nu=-0.00025\pm0.00013$. Overall, while none of the cases provides a conclusive detection, the combined datasets point
systematically toward a mild but persistent deviation from the standard $\Lambda$CDM scenario. The results for $\nu$ are also illustrated in
Fig.~\ref{fig:nu_posteriors}, where we show the respective marginalized posteriors indicating the GR limit.

Regarding the constraints on the Hubble constant, it also shows a systematic upward shift compared to the standard $\Lambda$CDM prediction from Planck,
$H_0 = (67.4 \pm 0.5)$ km\,s$^{-1}$\,Mpc$^{-1}$~\cite{Planck:2018vyg}. In the CMB-only analysis, we recover essentially the same result, $H_0 = (67.43 \pm 0.53)$ km\,s$^{-1}$\,Mpc$^{-1}$,
in excellent agreement with $\Lambda$CDM. Once the $H_0$ prior is included, however, the central value rises to $ ( 68.30 \pm 0.51)$ km\,s$^{-1}$\,Mpc$^{-1}$,
alleviating marginally the tension with the local measurement by about $1\sigma$. A similar effect is observed when BAO and SN Ia data are combined:
without the prior, we find $H_0 = (68.07 \pm 0.29)$ km\,s$^{-1}$\,Mpc$^{-1}$, and this increases further to $H_0 = (68.35 \pm 0.28)$ km\,s$^{-1}$\,Mpc$^{-1}$ with the $H_0$ prior. Although one can not affirm that
these shifts solve the Hubble tension, they systematically push the preferred value upward relative to the baseline Planck result, evidencing that
the model can accommodate higher $H_0$ values when additional low-redshift information is included. This effect is accompanied by correlated shifts in
other parameters. In particular, the scalar spectral index $n_s$ increases, moving closer to scale invariance when other probes are included.
Furthermore, the optical depth $\tau_\mathrm{reio}$ also increases with the inclusion of the $H_0$ prior, reflecting degeneracies in the parameter space.
Likewise, $\Omega_\Lambda$ grows slightly in the joint analyses, consistent with the higher expansion rate. Together, these results indicate that the
inclusion of late-time information tends to shift the cosmological parameter space in a coherent way, alleviating the $H_0$ tension while also affecting
early-Universe quantities such as $n_s$ and $\tau_\mathrm{reio}$.

\begin{table}[ht]
\centering
\resizebox{\textwidth}{!}{
\begin{tabular}{lcccc}
\hline\hline
 & CMB only
 & CMB + $H_0$ prior
 & CMB + BAO + SN Ia
 & CMB\,+\,BAO\,+\,SN\,Ia\,+\,$H_0$\,prior \\
\hline
$\log(10^{10} A_\mathrm{s})$
 & $3.037 \pm 0.014$
 & $3.045 \pm 0.014$
 & $3.042 \pm 0.014$
 & $3.046 \pm 0.016$ \\

$n_\mathrm{s}$
 & $0.9651 \pm 0.0046$
 & $0.9704 \pm 0.0045$
 & $0.9691 \pm 0.0036$
 & $0.9706 \pm 0.0036$ \\

$H_0$ [km\,s$^{-1}$\,Mpc$^{-1}$]
 & $67.43 \pm 0.53$
 & $68.30 \pm 0.51$
 & $68.07 \pm 0.29$
 & $68.35 \pm 0.28$ \\

$\Omega_\mathrm{b} h^2$
 & $0.02204^{+0.00015}_{-0.00018}$
 & $0.02209^{+0.00017}_{-0.00020}$
 & $0.02204^{+0.00016}_{-0.00019}$
 & $0.02209 \pm 0.00019$ \\

$\tau_\mathrm{reio}$
 & $0.0537 \pm 0.0072$
 & $0.0589^{+0.0069}_{-0.0080}$
 & $0.0575 \pm 0.0069$
 & $0.0600 \pm 0.0079$ \\

$\nu$
 & $-0.00014 \pm 0.00013$
 & $-0.00024 \pm 0.00014$
 & $-0.00023 \pm 0.00012$
 & $-0.00025 \pm 0.00013$ \\

$\Omega_\Lambda$
 & $0.6889 \pm 0.0072$
 & $0.7004 \pm 0.0067$
 & $0.6977 \pm 0.0037$
 & $0.7011 \pm 0.0036$ \\
\hline\hline
\end{tabular}}
\caption{Results of the statistical analysis. Each row lists the mean value and 68\% confidence interval of the corresponding cosmological parameter for the different dataset combinations.}
\label{tab:cosmo_results}
\end{table}

\begin{figure}[ht]
    \centering
    \includegraphics[width=0.49\textwidth]{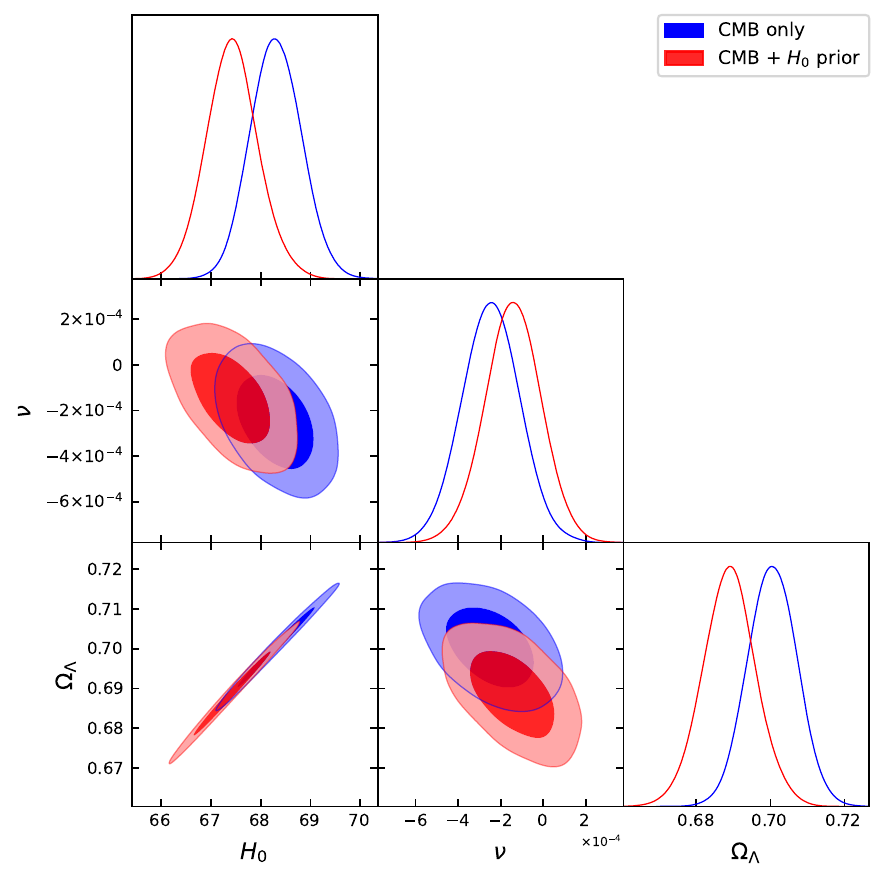}
    \includegraphics[width=0.49\textwidth]{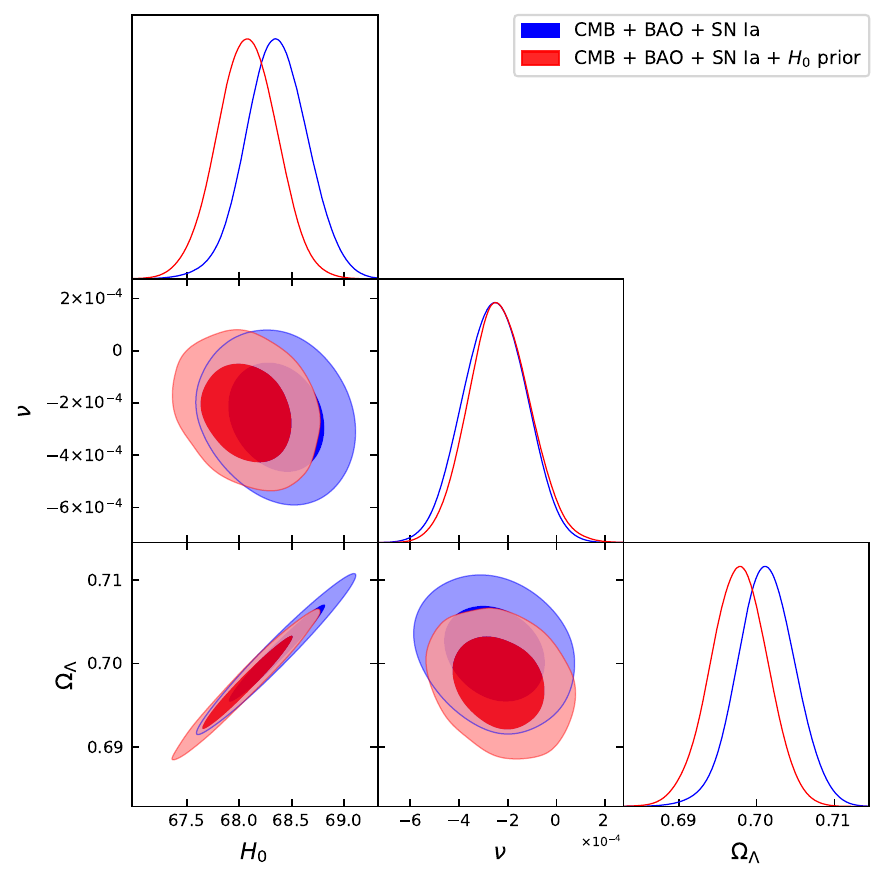}
    \caption{Corner plot for the most important set of cosmological parameters, namely $\{H_0, \Omega_\Lambda, \nu\}$. \textbf{Left panel: }results from CMB only and CMB + $H_0$ prior. \textbf{Right panel: }results from CMB + BAO + SN Ia and CMB + BAO + SN Ia + $H_0$ prior.}
    \label{fig:joint_constraints}
\end{figure}

The multidimensional results are also illustrated in Fig.~\ref{fig:joint_constraints},
where we show the corner plot with the 2D and 1D marginalized posteriors for the
set of parameters $\{H_0, \Omega_\Lambda, \nu\}$, which are the most relevant parameters
for our analysis. The figure clearly shows the correlation between $H_0$ and $\nu$,
with negative values of $\nu$ favoring higher $H_0$. The GR limit, $\nu=0$, is only
marginally compatible with the best fit, indicating a mild but persistent preference
for deviations from standard $\Lambda$CDM. The correlation between $H_0$ and
$\Omega_\Lambda$ is also evident, as expected from their roles in determining the
expansion history. Overall, the figure encapsulates the main findings of our analysis.

To quantify the impact on the Hubble tension, we estimate the discrepancy
between the inferred value of $H_0$ and the local SH0ES measurement in units of the combined
Gaussian uncertainty, namely
\begin{align}
T = \frac{|H_0^{\rm model}-H_0^{\rm SH0ES}|}
{\sqrt{\sigma_{\rm model}^2+\sigma_{\rm SH0ES}^2}} \, .
\end{align}
Using the CMB + BAO + SN Ia analysis, we obtain $T \simeq 3.6\sigma$.
Therefore, although the running framework does not fully resolve the
Hubble tension, it provides a modest but systematic reduction of the
discrepancy relative to the standard $\Lambda$CDM scenario.

\section{Conclusions}
\label{sec:conclusions}

Renormalization group corrections to the action of gravity may be a 
consequence of the quantum effects of massive matter fields. The 
energy scale of the Universe, in the cosmological context, can be 
associated with the Hubble rate $H$. In the low-energy regime, 
when the masses of quantum fields are much greater than the
magnitude of $H$, the quantum corrections may be quadratically
suppressed compared to the leading logarithmic corrections in the
high-energy regime (UV), but are still relevant because the
unsuppressed beta function for the cosmological constant density
$\rho_\La$ and for the inverse Newton constant $1/G$ are of the
fourth and second order in masses, respectively. The independent 
arguments based on the assumption of a standard quadratic decoupling 
\cite{AC}, on the covariance of effective action and on simple 
dimensional arguments converge to the universal form of the 
possible running of the dimensional constants (\ref{runCCmu0}) 
and (\ref{eq:runG}), with a single free parameter $\nu$. This 
quantity may be defined only from the observations. From the 
theoretical side, the value of this parameter depends on the largest 
masses of the underlying quantum theory of matter fields.

There is an important aspect of quantum corrections, which is worth 
an additional comment in this conclusions. It concerns the 
identification of the energy scale $\mu$ in the model with running 
$\rho_\La$ and $G$. In general, the most consistent way to introduce 
the running of  the gravitational and cosmological constants would 
be through the two possible identifications of scale: one associated 
with the cosmological background, which can be related to the 
Hubble rate $H$, and another at the level of perturbations 
\cite{Bertini:2024onw}. The freedom in choosing the latter may 
lead to distinct effective energy densities and, consequently, to 
different equations. However, it was established in the same work, 
that if  there is no energy exchange between matter and vacuum, 
only the first background identification is truly relevant. The 
conservation of the energy-momentum tensor ensures that the 
perturbation of $\mu$ does not appear in the linearized equations
\cite{Fabris2010}, so the ambiguity does not arise in this regime. 
Moreover, not every scalar associated with $\mu$ that reproduces 
$\mu = H$ preserves this property at the level of perturbations.  
This feature explains why, in the present model, it is not necessary 
to postulate two distinct scales. 

In conclusion of the theoretical part of our consideration, the 
model we deal with represents the unique approach to running 
late-epoch cosmology which is consistent with the covariance of 
the vacuum effective action and with all principles and presently
known sound results of quantum field theory in curved spacetime. 
This uniqueness concerns not only the behavior of the background
metric, but also cosmic perturbations. 

Comparing this framework with observational data, our statistical analysis based
on the most recent publicly available cosmological data provides a consistent picture
of the impact of this running framework. Across all dataset combinations,
the parameter $\nu$ is consistently found to be negative, with statistical significance
ranging from $1.1\sigma$ (CMB only) up to nearly $2\sigma$ when BAO and SN Ia data are included.
The $\Lambda$CDM limit ($\nu=0$) thus remains marginally consistent with the data,
but the results show a mild and persistent preference for deviations. Interestingly,
negative values of $\nu$ correlate with slightly larger values of the Hubble constant.
These shifts are accompanied by correlated changes in other parameters, such as a higher
spectral index, slightly closer to unity, and an increase in the reionization optical depth,
reflecting the underlying parameter degeneracies. These results are compatible with the
EDE-like behavior of the model at early times, which is well-known to alleviate the $H_0$
tension (see also~\cite{Poulin2023} for a review and further references).

Altogether, our results indicate that the scale-dependent cosmology provides a theoretically well-motivated
and phenomenologically viable extension of $\Lambda$CDM. While the statistical evidence for $\nu \neq 0$
is not yet conclusive, the consistent preference for negative values across different datasets,
together with the partial alleviation of the $H_0$ tension, highlights the potential of this framework
to capture relevant quantum effects in the cosmological dynamics. Future high-precision observations,
in particular from next-generation CMB and large-scale structure surveys, will be crucial to determine
whether the running of gravitational couplings leaves observable imprints, and to test the possible link
between cosmology and the heaviest mass scales of particle physics.

\section*{Acknowledgments}

MN thanks the Fundação de Amparo à Pesquisa e Inovação do Espírito Santo (FAPES) and the hospitality of the Federal University of Bahia.
RvM is supported by Fundação de Amparo à Pesquisa do Estado da Bahia (FAPESB) grant TO APP0039/2023 and CNPq (Conselho Nacional de Desenvolvimento Científico e Tecnológico, Brazil) under the grant 311114/2026-1.
I.Sh. is grateful to CNPq (Conselho Nacional de Desenvolvimento
Cient\'{i}fico e Tecnol\'{o}gico, Brazil)  for the partial support
under the grant 305122/2023-1.

\section*{Data availability Statement}

No new data were generated or analyzed in support of this research.

\appendix

\section{Scale-setting and energy-momentum tensor conservation}
\label{sec:apdx1}

In this section we show that the scale-dependent scenario
 developed in Sec.~\ref{sec22} is compatible with $\nabla_{\al}T^{\al\be} = 0$. Let us start with the scale setting procedure, which consists of establishing a relationship between $\mu$ and the physical information contained in $\rho$. By contracting Eq.~\eqref{eq:FE} with $u^{\al}u^{\be}$ and using the energy-momentum tensor \eqref{eq:EMT}, we obtain
\begin{align}
    \rho + \rho_\Lambda = \frac{1}{8\pi G}u^{\al}u^{\be}G_{\al\be}.\label{eq:scale-setting}
\end{align}
In this relationship, the parameters $\rho_\Lambda$ and $G$ depend on $\mu$ in the RG framework. Therefore, the connection between $\mu$ and $\rho$ must involve $u^{\al}u^{\be}G_{\al\be}$. This motivates the definition
\begin{align}
\mu^2 := \frac{1}{3}u^\al u^\be G_{\al\be}\,\label{eq:ss}
\end{align}

Let us move on to the conservation law. We follow the standard construction and consider that, since $u^\al u^\be g_{\al\be}=1$, any arbitrary vector $A^{\al}$ can be decomposed into components parallel and orthogonal to $u^{\al}$ as
$ A^{\be} = (u_\al A^{\al})u^{\be} + P^{\be}_{\;\;\al}A^\al$,
where $P^{\al\be} := u^{\al}u^{\be}-g^{\al\be}$ is a projector orthogonal to $u^{\al}$, i.e., $P^{\be}_{\,\,\,\al}u^{\al}=0$ and $P^{\al}_{\;\;\si}P^{\si}_{\;\;\be} = P^{\al}_{\;\;\be}$. With this, we can write
\begin{align}
    \nabla_{\al}T^{\al\be} = (u^\si \nabla_{\al}T^{\al}_{\,\,\,\si})u^{\be} + P^{\be}_{\,\,\,\si}\nabla_{\al}T^{\al\si}\,.
\end{align}
Therefore, $\nabla_{\al}T^{\al\be} =0$ if and only if
\begin{align}
u^\si \nabla_{\al}T^{\al}_{\,\,\,\si}&=0\,,\label{eq:cont1}\\
P^{\be}_{\,\,\,\si}\nabla_{\al}T^{\al\si}&=0\,.\label{eq:euler}
\end{align}
These equations are the relativistic versions of the continuity and Euler equations, respectively. The strategy is to show that these identities hold.

Taking the divergence of Eq.~\eqref{eq:FE}, the Bianchi identities directly imply that
\begin{align}
    \nabla_{\al}T^{\al}_{\;\;\be} = -\nabla_\be \rho_\Lambda - (T^{\al}_{\;\;\be} + \rho_\Lambda \delta^{\al}_{\be})G^{-1}\nabla_{\al}G\,.\label{eq:divT}
\end{align}
Contracting with $u^{\be}$, considering the energy-momentum tensor of a perfect fluid \eqref{eq:EMT} and taking into account that $\nabla_{\al}\rho_\Lambda = (\nabla_{\al}\mu)d\rho_\Lambda/d\mu$ and $\nabla_{\al}G = (\nabla_{\al}\mu)dG/d\mu$, we obtain
\begin{align}
    u^{\be}\nabla_{\al}T^{\al}_{\;\;\be} = -\left( \frac{d\rho_\Lambda}{d\mu} + \frac{\rho+\rho_\Lambda}{G}\frac{dG}{d\mu} \right) u^{\al}\nabla_{\al}\mu\,.
\end{align}
Using relation \eqref{eq:scale-setting} and \eqref{eq:ss}, we conclude that in order to obtain $u^{\be}\nabla_{\al}T^{\al}_{\;\;\be} = 0$ with $u^{\al}\nabla_{\al}\mu\neq0$, it must be that
\begin{align}
    \frac{d\rho_\Lambda}{d\mu} - \frac{3\mu^2}{8\pi}\frac{dG^{-1}}{d\mu}=0.\label{eq:consist1}
\end{align}

Let us now consider the equality \eqref{eq:euler}. Contracting \eqref{eq:divT} with $P^{\si\be}$ and using the field equation \eqref{eq:FE}, we get
\begin{align}
    P^{\be}_{\;\;\si}\nabla_{\al}T^{\al \si} = -P^{\be}_{\;\;\si}\left(\frac{d\rho_\Lambda}{d\mu}g^{\al\si} - \frac{G^{\al\si}}{8\pi}\frac{dG^{-1}}{d\mu} \right)\nabla_{\al}\mu\,.\label{eq:euler1}
\end{align}
Here is the key step that demonstrates the compatibility of the identification \eqref{eq:ss} with the conservation law. Using this identification along with \eqref{eq:consist1}, one gets
\begin{align}
    \nabla_{\al}\mu = \frac{1}{6\mu}\nabla_{\al}(u^{\si}u^{\rho}G_{\si\rho}) = \frac{4\pi}{3\mu}\nabla_{\al}[G(\rho+\rho_\Lambda)] = \frac{4\pi G}{3\mu}\nabla_{\al}\rho\,.
\end{align}
Then, substituting this result into \eqref{eq:euler1} and using \eqref{eq:consist1} to eliminate  $d\rho_\Lambda/d\mu$,  Eq.~\eqref{eq:euler1} becomes
\begin{align}
    P^{\be}_{\;\;\si}\nabla_{\al}T^{\al \si} = -P^{\be}_{\;\;\si}\big(3\mu^{2} g^{\al\si} - G^{\al\si}\big)\frac{G}{6\mu}\frac{dG^{-1}}{d\mu} \nabla_{\al}\rho\,.\label{eq:PnablaT}
\end{align}
For the right-hand side of this equation to vanish, it is necessary that $\big(3\mu^{2} g^{\al\si} - G^{\al\si}\big)\nabla_{\al}\rho \propto u^{\si}$. One way to ensure this proportionality is to consider a barotropic fluid without vorticity, i.e., $\omega_{\al\be}=P_{\al}^{\;\mu}P_{\be}^{\;\nu}\nabla_{[\mu}u_{\nu]}=0$. In this case, there exists locally a scalar potential $\theta(x^{\al})$ and a function $F(x^{\al})$ such that \cite{lichnerowicz1967}
\begin{align}
    u_\al = F\nabla_\al\theta\,,
\end{align}
with $F = (\nabla_{\al}\theta \nabla^{\al}\theta)^{-1/2}$ fixed by the normalization condition $u^{\al}u_{\al}=1$. In the barotropic case, the condition $\omega_{\al\be}=0$ is preserved along the dynamical evolution \cite{landau1987, anile1990}. Moreover, the continuity equation $u^{\si}\nabla_{\al}T^{\al}_{\;\;\si} = u^{\al}\nabla_{\al}\rho + \nabla_{\al}u^{\al}(\rho+p)=0$ implies that $\rho = \rho(\theta)$. Consequently,
\begin{align}
    \nabla_{\al}\rho = \rho' \nabla_{\al}\theta = \rho' F^{-1}u_{\al}\,,
\end{align}
where $\rho' = d\rho/d\theta$. Substituting this relation into \eqref{eq:PnablaT} we obtain
\begin{align}
    P^{\be}_{\;\;\si}\nabla_{\al}T^{\al \si} = -P^{\be}_{\;\;\si}\big(3\mu^{2} u^{\si} - G^{\al\si}u_{\al}\big)\frac{G\rho'}{6\mu F}\frac{dG^{-1}}{d\mu}\,.
\end{align}
Since $P^{\be}_{\;\;\si}u^{\si}=0$, the first term in parentheses vanishes automatically. The second term can be rewritten as
\begin{align}
    P^{\be}_{\;\;\si}G^{\al\si}u_{\al} = (u^{\be}u_{\si}-\delta^{\be}_{\si})G^{\al\si}u_{\al} = 3\mu^{2}u^{\be}-u_\al G^{\al\be} \,,
\end{align}
which vanishes only if $3\mu^{2}u^{\be} = u_\al G^{\al\be}$.
This condition is indeed satisfied, as both sides reduce to $3\mu^2$ when contracted with $u^\be$, consistently with the definition \eqref{eq:ss}.


\end{document}